\documentstyle[amsfonts,12pt]{article}
\newcommand{\be}{\begin{equation}}
\newcommand{\ee}{\end{equation}}
\newcommand{\bea}{\begin{eqnarray}}
\newcommand{\eea}{\end{eqnarray}}
\newcommand{\ba}{\begin{array}}
\newcommand{\ea}{\end{array}}

\def\bbox{{\,\lower0.9pt\vbox{\hrule \hbox{\vrule height 0.2 cm
\hskip 0.2 cm \vrule height 0.2 cm}\hrule}\,}}
\newcommand{\dsl}{\pa \kern-0.5em /}

\newcommand{\nn}{\nonumber \\}

\font\mybb=msbm10 at 10pt
\def\bb#1{\hbox{\mybb#1}}

\begin{document}



\begin{titlepage}
\rightline{ }
\rightline{\tt }
\begin{flushright}
\end{flushright}

\vfill

\begin{center}
\baselineskip=16pt
{\Large\bf  Super Chern-Simons Quantum Mechanics{$^\star$}}
\vskip 0.3cm
{\large {\sl }}
\vskip 10.mm
{\bf  Luca Mezincescu$^{1}$  }
\vskip 1cm
 
{\small
Department of Physics,\\
University of Miami,\\
Coral Gables, FL 33124, USA\\
}
 \end{center}
\vfill

\par
\begin{center}
{\bf ABSTRACT}
\end{center}
\begin{quote}

The Super Chern-Simons mechanics, and quantum mechanics of a particle, on the  coset  super-manifolds $SU(2|1)/ U(2)$ and   $SU(2|1)/ {\left[U(1)\times U(1)\right]}$, is considered. Within a convenient quantization procedure  the well known Chern-Simons mechanics on $SU(2)/U(1)$  is reviewed, and then it is shown how the fuzzy supergeometries arise. A brief discussion of the super-sphere is also included.  

\vfill
\vfill
\vfill

$^\star$ {  To appear in Proceedings                                
 of the International Workshop "Supersymmetries and Quantum                     
 Symmetries" (SQS'03, 24-29 July, 2003), Dubna, Russian Federation. }

\vfill
\hrule width 5.cm
\vskip 2.mm
{\small
\noindent $^1$ mezincescu@server.physics.miami.edu\\
\\ }
\end{quote}
\end{titlepage}

\setcounter{equation}{0}
\section{Introduction}

The actions corresponding to anti-commuting degrees of freedom  contain only first order time derivatives of the corresponding fields. This  suggests that such actions  may be somehow connected with with the pull-back of a one-form. But if the action is to be a pull-back of a one form  then the base space must be one dimensional, therefore the corresponding action generated by a one form should be that of a particle moving on a super-manifold.  Moreover such a one form may be chosen  to be the connection form on some super-manifold, associating a certain  ``dynamics''  with the given ``affine geometry''. One  then deals with the Super Chern-Simons mechanics. Studies of the Chern-Simons mechanics and its  world line super-symmetric generalization have been performed before \cite{DJT,HT}.  In what follows some recent work \cite{IMPT,IMT} on particle motion on supermanifolds, whose actions are such Super Chern-Simons terms will be described. In fact for simplicity only various cosets of $SU(2|1)$ will be considered, as they illustrate well, the generic case treated, which is the super Kahler manifold $CP^{\left(n|m\right)}$.

The bosonic Chern-Simons actions are prototype actions which lead to fuzzy geometry \cite{ madore,GS, TV, ABIY,BDL,AMMZ}, under quantization. Their counterpart the Super Chern-Simons actions will also be shown to lead to the fuzzy supergeometry (see also \cite{martin,BS,FL,KPT,dBGvN,OV,seiberg}). The quantization procedure will be first described for the simple case of the  Chern-Simons action on the sphere, then the same procedure will be applied to the Super Chern-Simons actions corresponding to the $U(1)$ connection associated with the supermanifolds $SU(2|1)/U(2)$  and $SU(2|1)/U(1|1)$.  In the process the definition of the supersphere will be also touched upon. 

\section{Chern-Simons Quantum Mechanics on  $\bf  S^2$}

Let  $z$ be the complex coordinate on the sphere defined by stereographic projection to the complex plane, and let $z(t)$ be the classical position of a charged particle on the sphere. The Chern-Simons mechanics for this particle is defined by the Lagrangian 
\be\label{lagr}
 L =  N\left( {\dot z }{A_z} + {\dot {\bar z }}{A_{\bar z}}\right),
\ee 
where  $2N$ is an integer (at the quantum level), and
\be\label{connection}
A_z = -i\partial_z {\ln K_0}, \qquad A_{\bar z} = i\partial_{\bar z} \ln K_0, 
\ee
are the complex components of the $U(1)$ connection on the sphere, (which can also be viewed as the gauge potential due to a unit charge monopole at the centre of the sphere) with
\be
 K_0 =\left(1+ z\bar z\right).
 \ee
 This Lagrangian is invariant  (up  to a total derivative) under the $SU(2)$ isometry group of the sphere (in fact it is invariant under the infinite dimensional group of simplectic diffeomorphisms on the sphere, which is why these models are topological), the infinitesimal $SU(2)$ transformations are obtained from the closure  of
\be\label{isometry}
\delta z = \varepsilon + \bar \varepsilon z^2 , \qquad \delta \bar z = \bar\varepsilon + \varepsilon \bar z^2,
\ee
for complex parameter $\varepsilon$, under which:
\be
\delta_\varepsilon K_0 =
\left(\bar\varepsilon z +\varepsilon {\bar z}\right) K_0 ,
\ee
and  (\ref{lagr})  transforms by a total derivative. One could pass to the quantum theory by the usual  Dirac method prescription,  but  another method will be used with some advantages that will hopefully become apparent in what follows. With the standard definition of the canonical momenta, the two second class constraints of this model can be expressed\footnote{One uses  the symbol $\approx$ to denote weak equality in the sense of Dirac.} as the complex conjugate pair of constraints
$\varphi_z \approx 0$ and $\varphi_{\bar z} \approx 0$, where
\be\label{constraints}
\varphi_z = p_z - NA_z , \qquad  \varphi_{\bar z} = p_{\bar z} -NA_{\bar z}. 
\ee
One now quantizes , as if there were no constraints, by setting
\be\label{quantization}
p_z = {1\over i} {\partial \over {\partial z}}\quad \quad, \quad\quad  p_{\bar z} =
{1\over i} {\partial \over {\partial \bar z}}, 
\ee
and  then imposes the physical state condition 
\be\label{physical}
\varphi_{\bar z} {\Psi \left( z ,{ \bar z} \right)}_{phys}    = 0, 
\ee
which has the solution 
\be\label{wavefunction}
{\Psi \left( z ,{ \bar z} \right)}_{phys} = {\left( 1 + z{\bar z} \right)}^{-N} {\Phi \left( z \right)}.
\ee
for holomorphic function $\Phi(z)$. The $SU(2)$-invariant inner product of two wave-functions $\Psi$ and $\Omega$ corresponding, respectively, to the holomorphic functions $\Phi(z)$ and $\Upsilon(z)$ is \be\label{inner}
\left( \Psi , \Omega\right) = \int \!{dzd\bar z\over \left(1 + z\bar z\right)^2}{\bar \Psi}\left({\bar z}, z\right)\,\Omega\left( z, {\bar z}\right)\
= \int \! {{{dzd{\bar z}}{}\over {{\left( 1 + z\bar z\right)^{2\left(N+1\right)}}}}\,
\bar {\Phi}\left(\bar z\right) \Upsilon\left(z\right)}. 
\ee
Normalizability of  $\Psi$ and $\Omega$ in this inner product requires $\Phi(z)$ and $\Upsilon(z)$ to be polynomials in $z$ of maximum degree $2N$. Correspondingly the physical Hilbert space is  $2N+1$
dimensional.

For fermionic constraints, this alternative method of dealing with
second-class constraints can be traced back to the 1976 papers of
Casalbuoni \cite{casal} and papers in the early 1980s of Azc\'arraga et al.
\cite{cowork,cowork2} and Lusanna \cite{LL}. A clear statement of it can be
found, again for fermionic constraints, in a 1986 paper of de
Azc{\'a}rraga and Lukierski \cite{AL}, who called it `Gupta-Bleuler'
quantization by analogy with the procedure of that name for covariant
quantization of electrodynamics \footnote{Noting that the Lorentz gauge
condition cannot be consistently imposed as a physical state condition,
Gupta and Bleuler suggested that it be separated into its positive and
negative frequency parts (of which $\varphi$ and $\bar\varphi$ are
analogs) and that the positive frequency part be imposed as the physical
state condition.}. It was also called Gupta-Bleuler quantization in
the 1991 book of Balachandran et al.\cite{Bbook}, where it is explained
for particle mechanics models with bosonic constraints. The justification
for this method sketched above arose in independent work on
general models with bosonic second-class constraints that can be separated
into two sets of {\it real} constraints, each in involution \cite{MR,HM}.
In this context the method has become known as the method of
`gauge-unfixing'.

As the Lagrangian  (\ref{lagr}) is invariant (up  to a total time derivative)  under the $SU(2)$ isometry group  of the sphere, there should be an action of this group induced on the physical Hilbert space. Allowing for operator ordering ambiguities, the  Noether charge operators that generate the  infinitesimal  transformations (\ref{isometry}) are
\be
J _+= i\left( P_{\bar z} + z^{2}P_{z}  + i\alpha z\right) , \qquad 
J_-  = i\left( P_{z} +{ \bar z}^{2}P_{\bar z}  -i\beta{ \bar z}\right)
\ee
for some constants $\alpha$ and $\beta$, leading to the transformation property of $\Psi$,
\be
\delta \Psi  =  \left(\epsilon J_{+} +  {\bar \epsilon} J_{-}\right)\Psi .
\ee
A necessary condition for these charges to take physical states into physical states is that they commute (weakly) with $\varphi_{\bar z} $, and this fixes $\beta=N$. The constant $\alpha$ remains undetermined by this requirement, however when constructing a representation of $SU(2)$ one must demand that $N+ \alpha $ is an integer and than
the representation is  $N+ \alpha +1$ dimensional. It is therefore possible to choose a natural value $\alpha = \beta =  N^{\prime}$, with $2N^{\prime}=N + \alpha$, without loss of generality.  Then: 
\be
J_- = -J_+^\dagger,
\ee
with respect to the inner product (\ref{inner}). Thus, 
\be\label{Jays}
J _+=   \left(  \partial_{\bar z} + z^2\partial_z\right) - Nz, \qquad
J_- = \left(\partial_z + \bar z^2\partial_{\bar z}\right) + N\bar z.
\ee
Note that
\be
J_\pm \Psi = {\left( 1 + z{\bar z} \right)}^{-N} j_\pm {\Phi \left( z \right)},
\ee
where
\be\label{smallcasejays}
j_-= \partial_z ,\qquad j_+ = z^2\partial_z -2Nz
\ee
are the charge operators acting on holomorphic functions. These have the commutator
\be
[j_-,j_+] = 2j_3, \qquad j_3 = z\partial_z -N,
\ee
and $(j_-,j_+,j_3)$ span the Lie algebra of $SU(2)$. Monomials in $z$ are eigenfunctions of $j_3$, with eigenvalues that range between $-N$ for constant $\Phi$ and $+N$ for $\Phi \propto z^{2N}$. The polynomials of maximal degree $2N$ therefore span the $2N+1$ irrep of $SU(2)$, and hence the $2N+1$-dimensional Hilbert space is a carrier space for this irrep. 
The operators $J_\pm$ generate translations on the sphere, and act on the physical Hilbert space, so they are naturally identified with the momentum operators of the quantum theory. But what are the position operators corresponding to the classical position variable $z$ and its complex conjugate $\bar z$? On general grounds one may expect some complications for the definition of the position operators due to the fact that the naive definition does not lead to operators with tensorial properties. 
Also,  $\bar z \Psi(z)$ is clearly not  a physical state (due to the fact that $\bar z$ does not commute with $\varphi_{\bar z} $) and if $\Phi$ is a polynomial of maximal degree then $z\Psi$ is not a physical state either (despite the fact that $z$ commutes with  $\varphi_{\bar z} $). Thus, the `naive' position space operators do not act on the physical Hilbert space. This could have been anticipated from the fact that $z$ and $\bar z$ commute whereas physical position operators are non-commutative in first order systems \cite{DJT}. However, this non-commutativity must disappear in the $N\rightarrow\infty$ limit because  this limit is in fact the semiclassical limit of this  problem, Therefore  the position operator $z$  takes the form:
\be
z_{op} = z + {\cal O}(1/N), 
\ee
where $z$  is the naive position operator. The ${\cal O}(1/N)$ corrections must be such as to ensure that $z_{op}$ and ${\bar z}_{op}$, act on the physical Hilbert space. There is a unique solution to this problem, with a minimal number of derivative operators, and the result is
\be
z_{op} = z - {1\over N}\left(\partial_{\bar z} + z^2{\partial_z}\right), \qquad
{\bar z}_{op} = \bar z + {1\over N}\left( \partial_{z}+ {\bar z}^2 \partial_{\bar z}\right). 
\ee
Comparison with (\ref{Jays}) shows that
\be\label{coord}
z_{op} = -{1\over N}J_+, \qquad \bar z_{op} = {1\over N}J_-. 
\ee
That is in  terms of (\ref{smallcasejays}),  one  just discards the term which does not lead to a normalizable state upon the action of naive $z$. The position operators are proportional to the momentum operators, and thus span the algebra of the $SU(2)$ isometry group of the sphere. The proportionality between the position operators and the corresponding generators of $SU(2)$ is the reason one calls the corresponding manifold ``fuzzy''. It is clear that one  can exhibit the fact that the position operators correspond to the quantization of certain observables of the classical theory (which in
fact corresponds to a reshuffling of classical phase space variables)
 \bea
{z}_{op}  =   z -{i\over N} \left[{ p}_{\bar z}+ z^2
p_z\right]  =  {2z \over \left( 1 + z\bar z\right)} -  {i\over N}\left( {z^2}\varphi_z+ {\varphi}_{\bar z}\right)\approx
{2z \over 1 + \bar z\cdot z}\,.
\eea
\bea
{\bar z}_{op}  =   {\bar z} +{i\over N} \left[{ p}_{ z}+ {\bar z}^2
{ p}_{\bar z}\right]  =  {2{\bar z} \over \left( 1 + z\bar z\right)} + {i\over N}\left( {{\bar z}^2}\varphi_{\bar z}+ {\varphi}_{ z}\right)\approx
{2{\bar z} \over 1 + \bar z\cdot z}\,.
 \eea
The classical part of the operators $z, \bar z$ are in fact the conveniently redefined $x, y$ components of the position vector on the sphere as embedded in the three dimensional space. These classical position operators can be shown to obey Poisson brackets which close to the classical $SU(2)$ algebra. By defining the position operators in this model, one is therefore lead  to  the natural vector coordinate appearing in the problem. Here a limited set set of observables has been considered, it is clear however that once one knows the quantum system, which is a spin $N$ system,  one can introduce additional observables corresponding to the number of hermitian operators which can be built on the corresponding space. Even if the Hamiltonian for the system is absent, the requirement  of the existence of the scalar product restricts the space of states as well as leads to a correct definition of the observables for this system. 

\setcounter{equation}{0}
\section{Odd Coset Quantum Mechanics}

In what follows  the above described procedure, will be used to treat systems with anticommuting variables.  The simplest  purely odd supermanifold is  $SU(2|1)/U(2)$  ( $SU(2|1)$ superalgebra also has an involution). $SU(2|1)$ superalgebra consist of a even part, which is  $U(2)$,  and the odd sector which is made of two complex spinors under $SU(2)$ conjugate to each other, and of 
opposite ``baryon number'' corresponding to the $U(1)$.  The superflag manifold $SU(2|1)/U(2)$ has a  ``complex'' structure like the sphere, and it is purely odd. It is easy to show that a simple generalization of the transformation laws (\ref{isometry}), is:
\be
\label{antransf}
\delta_\epsilon \xi^i = \epsilon^i + \bar \epsilon\cdot \xi\, \xi^i\,,
\ee
where $\epsilon^i, \left(i= 1,2\right)$ are the two odd parameters corresponding to the odd transformations and $\xi^i$ are ``local'' anticommuting coordinates on the superflag. These  
transformations close to the superalgebra $SU(2|1)$, and one can guess the super Kahler like 
connection:
\bea
{A} = i {\left(d\xi \cdot \partial_{\xi} - d{\bar \xi} \cdot \partial_{\bar \xi}\right)} \ln K_1 
\eea
with:
\be
K_{1} = 1 + {\bar \xi}_{1} \xi^{1} +  {\bar \xi}_{2} \xi^{2},
\ee
where  $K_1$ transforms under (\ref{antransf}) as:
\be
\delta_\epsilon K_1 =
\left(\bar\epsilon\cdot\xi -\epsilon\cdot\bar\xi\right) K_1.
\ee

The corresponding Lagrangian is:
\be
\label{nonlinlag}
L =  {i\,\gamma\over 2} \left[1 + \bar\xi\cdot \xi\right]^{-1} \bar\xi\cdot \dot\xi\, + c.c.
\ee
and with the definition of the  odd canonical conjugate momentum:
\be
\pi_i = i {\partial L \over \partial {\dot {\xi}}^i}\quad \quad , \quad \quad {\bar \pi}^i = i {\partial L \over \partial {{\dot {\bar{\xi}}}}_ i}
\ee
leads to the second class constraints:
\be\label{defvarphi}
\varphi_i = \pi_i - {\gamma \over2}\left[1 + \bar\xi\cdot\xi\right]^{-1}\bar\xi_i\, \quad,\quad {\bar \varphi}^i = {\bar \pi}^i - {\gamma \over2}\left[1 + \bar\xi\cdot\xi\right]^{-1}\xi^i\,
\ee
Using the method described in the previous section, one obtains the larger phase space where the 
odd quantized  momenta are given by:
\be
\pi_i = {\partial\over\partial\xi^i}\, \quad,\quad
\bar\pi^i = {\partial\over\partial\bar \xi_i}\,.\label{conjugacy}
\ee
To take the constraints into account it must be required  that physical states be
annihilated by the operators $\varphi_i$,  this is equivalent to the
`analyticity' conditions\footnote{This is analogous to the chirality
condition on 4D chiral superfields. which arises in a similar way from
analytic quantization of the 4D superparticle \cite{Frydr,Dima}.}, or equivalently, with the condition that the corresponding covariant derivative be vanishing:
\be
{\partial\Psi \over \partial \xi^i} = {\gamma\over2}\left[1 +
\bar\xi\cdot\xi\right]^{-1}\bar\xi_i\Psi\, ,
\qquad i=1,2\,,
\label{cond}
\ee
on wave-functions $\Psi(\{\xi\},\{\bar\xi\})$. These conditions have the
solution
\be
\label{solution2}
\Psi = \left[1+\bar\xi\cdot\xi\right]^{-{\gamma\over 2}} \Phi
\ee
for anti-analytic $\Phi$, which has the expansion
\be
\Phi = a + \bar\xi_i b^i  + 
\bar\xi_{1} \bar\xi_{2}\ c\,.\label{expansion}
\ee
In this space an action of  $SU(2|1)$ should be introduced. The quantum version of the classical Noether generators corresponding to the transformation (\ref{antransf}) is:
\bea
\hat S_i &=& {\partial\over\partial\xi^i} + {{\alpha}\over2}\bar\xi_i -
\bar\xi_i
\left(\bar\xi \cdot {\partial \over \partial\bar \xi}\right), \nn
\hat {\bar S}{}^i &=& {\partial\over\partial\bar\xi_i} +
{{\beta}\over2}\xi^i +
\xi^i{\left(\xi\cdot {\partial \over \partial \xi}\right)}. \label{quantS}
\eea
The coefficients $\alpha$ and $\beta$ are undetermined at this stage due to the quantum ordering
ambiguities. The quantum operators $\hat S_i$ and $\hat {\bar S}{}^i$ must be however weakly commuting with the constraints (\ref{cond}), and this fixes $\beta = \gamma$ in (\ref{quantS}), in fact , one can also choose, with one notable exception, $\alpha = \beta$.  The action of these operators on the subspace of anti-analytic  superfields can be easily deduced from:
\be
\delta_\epsilon{\Psi}  \equiv -\left(\bar\epsilon\cdot \hat{\bar S} +
\epsilon\cdot{ \hat S}\right)\Psi =
\left[1+\bar\xi\cdot\xi\right]^{-{\gamma\over 2}}\delta_\epsilon{\Phi} .
\ee
where $\delta_\epsilon{\Phi} $ is given by:
\be\label{trans}
\delta_\epsilon\Phi =
-\left[\gamma(\epsilon\cdot\bar\xi) + \bar\epsilon\cdot
{\partial\over\partial\bar\xi}
- \left(\epsilon\cdot
\bar\xi\right) \bar\xi\cdot
{\partial\over\partial\bar\xi}\right]\Phi
\ee
For component fields in the expansion (\ref{expansion}) this
transformation implies:
\bea\label{transcomp}
&& \delta_{\epsilon}a = - \bar\epsilon_i b^i\,, \quad
\delta_{\epsilon}b^i = \gamma\epsilon^i a + \bar\epsilon_j
\varepsilon^{ij}c\;,\\
&& \delta_{\epsilon} c = (\gamma-1)\epsilon^{i}\varepsilon_{ij} b^{(j) }\;
\eea
where $\varepsilon^{12} = -\varepsilon_{12} = 1$, is the corresponding totally antisymmetric symbol.
One can see that for the special values of $\gamma = 0,1$ the corresponding transformations are reducible (but not totally reducible),  while for the generic values of $\gamma$ they are irreducible. In fact the case $\gamma = 0$, corresponds to the zero action, however it can be obtained from a non zero action using the arbitrariness in the definition of the supercharges mentioned before. To obtain the space of states however  an invariant inner product must be introduced. This can be accomplished by introducing the $SU(2|1)$ invariant measure:
\be
\int \!d\mu = \int \!d\mu_0 \left[1 + \bar\xi\cdot \xi\right]~,
\ee
where
\be
\int d\mu_0 = \prod_i {\partial\over\partial\bar\xi_i}
{\partial\over\partial\xi^i}\,.
\ee
Then the following bilinear form is $SU(2|1)$ invariant
\be
||\Psi_{(\gamma)}||^2=
\int\! d\mu \left[1 + \bar\xi\cdot\xi\right]^{-\gamma}
|\Phi_{(\gamma)}|^2\,.
\label{norm2}
\ee
Within this ``norm'' one can see that for generic $\gamma$'s the space of states will consist  of four dimensional irreducible multiplet of $SU(2|1)$, corresponding to $q ={1\over 2}$,  \cite{NRS}.
For the special values $\gamma = 0,1$ the space of states consists of a degenerate representation of 
$SU(2|1)$ \cite{NRS} and respectively a singlet. In this case therefore the wave vector contains zero norm states.  $\gamma$ is the $U(1)$
``baryon'' charge in the $SU(2|1)$ algebra, it is not quantized like it usually happens for the WZ-like terms, however simplifications associated with integer values nevertheless appear for certain
integer values of it. 

In the special case when $\gamma =1$ it is possible to introduce an alternative invariant norm, if one forces the singlet to vanish by imposing the covariant condition:
\be
c = 0 \quad \Leftrightarrow \quad
\frac{\;\;\partial^2 \tilde{\Phi}_{(1)}}{\partial \bar\xi_{i}\partial
\bar\xi_k} = 0~.
\ee
The alternative norm is
\bea
||\tilde{\Phi}_{(1)}||^2 &=& - \int d\mu_0\,\ln (1 +\bar\xi\cdot \xi)\,
\tilde{\bar\Phi}_{(1)}\tilde\Phi_{(1)} \nn
&=& |a|^2 + \bar b_i b^i \label{norm3}
\eea
and it projects again a degenerate representation of $SU(2|1)$. 

A discussion, similar to that in the previous section, of the position operators can be performed  with the analogous result that  the naive odd operators must be redefined and the newly defined operators are just the odd generators of the $SU(2|1)$, while the classical odd coordinates are the ``projective '' coordinates:
\begin{equation}
W^i = \gamma \left(1+ {\bar \xi}\cdot \xi\right)^{-1} \xi^i.
\end{equation} 

\setcounter{equation}{0}
\section{Supersphere}

In what follows Chern-Simons mechanics with both even and odd coordinates will be considered.  One  therefore chooses other cosets of $SU(2|1)$. Any coset of  $SU(2|1)$ whose body is $S^2$ can be viewed as a graded generalization of the sphere. By considering an appropriate  parametrization of $SU(2|1)/\left[U(1)\times U(1)\right]$ it will be  shown  that one is lead to a minimal extension of the sphere by spinor coordinates. Indeed it  can be shown  \cite{IMTP}  that the following transformations:
\bea\label{transf}
\delta  z &=&  \varepsilon + {\bar \varepsilon}  z^2 - \left(\bar\epsilon_2 +
z\bar\epsilon_1\right)\left(\xi^1 -z\xi^2\right) \nn
\delta \xi^1 &=& {\varepsilon} \xi^2 + \epsilon^1 + \left(\bar\epsilon\cdot
\xi\right)\xi^1 \nn
  \delta \xi^2 &=& -{\bar \varepsilon}\xi^1 + \epsilon^2 + 
\left(\bar\epsilon\cdot\xi\right)
\xi^2
\eea
close to the algebra of $SU(2|1)$, these transformations correspond to the isometries of the $SU(2|1)/\left[U(1)\times U(1)\right]$ supermanifold which is a graded generalization of $S^2$. Then, one  can introduce an action of $SU(2|1)$ on a subsupermanifold of the above supermanifold. Consider the local coordinates $z$ and $\xi = \xi^1 -z\xi^2 $, they transform among themselves under the above transformations:
\bea\label{superstr}
\delta  z &=&  \varepsilon + {\bar \varepsilon}  z^2 - \left(\bar\epsilon_2 +
z\bar\epsilon_1\right)\xi \nn
\delta \xi &=&\left({ \bar \varepsilon}z\right)\xi  + \epsilon^1- {\epsilon}^2 z  \, 
\eea
What happens is that $z$ and $\xi$ transform linearly, among themselves, under the $\epsilon^2$ transformation which therefore passes into the stability group. The stability group consists now of the corresponding odd generators, and of the two $U(1)$'s in the $SU(2|1)$. Therefore it is the $U(1|1)$. The two  $U(1)$'s can be redefined, so that one of them commutes with the odd generators in the stability group, while the other $U(1)$ appears in the anticommutator of the corresponding odd generators. One therefore has only one abelian connection which transforms by a total derivative under the  motions on the supermanifold,
which  are those corresponding to $\varepsilon$ and $\epsilon^1$.  This supermanifold is   an extension of the sphere which maintains its complex structure and it can be easily shown that it is a homogenous  symmetric space. It has been called \cite{IMT}  the supersphere
\footnote{The term
`supersphere' has been used previously for the coset superspace
$UOSp(1|2)/U(1)$ \cite{GKP,GR2,BKR,HIU,IU}. Although this superspace is
often stated to have real dimension $(2|2)$, its `reality' is defined with
respect to a `pseudoconjugation'; see e.g. \cite{AzA1} for details. With
respect to standard complex conjugation, it actually has real dimension
$(2|4)$ (that is the same dimension as the $SU(2|1)/U(1)\times U(1)$ example which however is not a symmetric space) since spinors of $USp(2)\cong SU(2)$ span a vector space of
dimension $4$ over the reals.}. It will be  shown
that the `Hilbert' space of a particle on a supersphere at fuzziness
level $2N$ is a degenerate irrep of $SU(2|1)$, $q = N$ \cite{NRS} that decomposes
with respect to $SU(2)$ into a supermultiplet of $SU(2)$ spins
$(N-{1\over2},N)$. 

One  proceeds in a way similar to the previous section to find the connection forms:
\be
{A} =- i \left (dZ^M \partial_M    - d{\bar Z}_M \partial^M \right)\ln K_2  ,
\ee
with ($Z^M = z, \xi$) and ($ {\bar Z}_M = {\bar z}, {\bar \xi}$), and 
\be
K_{2} = 1 + {\bar z} z +  \xi {\bar \xi} ,
\ee
where  $K_2$ transforms under (\ref{superstr}) as:
\be
\delta_\epsilon K_2 =
\left({\bar\varepsilon} z + \varepsilon {\bar  z} +\epsilon^1 \bar\xi  - {\bar \epsilon}^1 \xi \right) K_2,
\ee
The Lagrangian is then:
\begin{equation}\label{CQSMlag}
L = -iN \left(1+  Z\cdot \bar Z\right)^{-1} \dot Z \cdot \bar Z +c.c.
\end{equation}
and leads to the physical wave function:
\begin{equation}
\Psi = \left[1 +  Z\cdot  \bar Z\right]^{-N}\Phi(Z)\,
\end{equation}
for holomorphic superfield $\Phi$, with the corresponding ``norm''
\bea\label{gennorm1}
||\Psi||^2   =  \int d\mu_0 \left[1 +  Z \cdot \bar Z\right]^{-\left(2N +1\right)}|\Phi|^2 
\eea
where (allowing for an arbitrary normalization factor ${\cal N}$)
\bea
\int d\mu_0 = {\cal N} \int \int dz d\bar z  {\partial\over
\partial \bar\xi} {\partial\over \partial \xi}\, .
\eea 
Therefore normalizability of $\Psi$ requires $\Phi$ to be a polynomial in $Z$ of maximum degree
$2N$, that is in this case as expected the WZ term is quantized, and $\Phi$ has the expansion:
\begin{equation}
\Phi = \phi_0(z) + \xi \phi_1(z)
\end{equation}
where $\phi_0$ and $\phi_1$ are two holomorphic functions of opposite
Grassmann parity. After performing the Berezin integrals over
$\xi$ and $\bar\xi$, one finds that
\begin{equation}
||\Psi||^2 \propto \left[ \int_{S^2} {dz d\bar z \over (1+ \bar z
z)^{2N+1}}\, |\phi_1|^2 + (1+2N) \int_{S^2} {dz d\bar z \over (1+ \bar z
z)^{2N+2}}\, |\phi_0|^2 \right].
\end{equation}
Normalizability of the second integral implies
that $\phi_0(z)$ is a polynomial of maximum degree $2N$ and hence that its
coefficients transform as spin $N$ under $SU(2)$, while normalizability of
the first integral implies that $\phi_1(z)$ is a polynomial in $z$ of
maximum degree $(2N-1)$, and hence that its coefficients transform as spin
$N-{1\over2}$ under $SU(2)$. If the spin-statistics connection is to be respected \footnote{This is not a mathematical necessity here because
the spins are non-relativistic.} then $\Phi$ should be chosen to have
Grassmann parity $(-1)^{2N}$.  Then  the `Hilbert' space that is a
supermultiplet with spins $\left(N -{1\over 2}, N\right)$
carrying a ${\bf 2N}\oplus ({\bf 2N+1})$ representation
of $SU(2)$; this is the decomposition into $SU(2)$ irreps of the
`degenerate' irrep of $SU(2|1)$ of total dimension $4N+1$ ($q = N$ in the notation of  \cite{NRS}).
As in the preceding sections the corresponding position operators become proportional with the 
corresponding charges of $SU(2|1)$.

In conclusion a very simple setting for the study of Super Chern-Simons quantum mechanics has been outlined. It produces the starting point for the study of fuzzy super-geometries, by generating  the corresponding ``fuzzy'' commutation relations among the coordinates,  within a  convenient quantization procedure. This approach is a useful companion to the methods of geometric quantization on compact simplectic manifolds. It
illustrates  within an extended phase space  associated with  Chern-Simons action, the quantization, of the Poisson brackets associated with these manifolds. The existence of the  Chern-Simons action, implies that of a non-degenerate two form, which in turn guarantees the existence of the  Poisson bracket mentioned before and vice versa, in order to construct a Poisson bracket one needs a non degenerate closed two form  (see e.g. \cite{DNF}) , whose corresponding potential one form generates the Chern-Simons action. 
\bigskip
\noindent

{\bf Acknowledgements.} Some unpublished material related to \cite{IMPT, IMT, IMTP}, has been used
while preparing this contribution, I am grateful to my coauthors on these references for granting me their permission to do that. I wish to thank the theory group at JINR for hospitality and partial support. Some of the results presented here have also been part of a talk at the 
 {\sl Workshop on Branes and Generalized Dynamics} (Argonne, October
20-24, 2003).  This work was supported in part by the National Science Foundation under 
grant PHY-9870101.

\end{document}